\documentclass[11pt,twocolumn]{article}
\usepackage[dvips]{graphicx}
\usepackage{amsmath}

\begin{document}


\title{Excitonic order at strong-coupling: pseudo-spins, doping, and
  ferromagnetism}
\author{Leon Balents \\  Physics Department, University of
  California, Santa Barbara, CA 93106}

\date{\today}

\maketitle

\begin{abstract}
  A tight binding model is introduced to describe the strong
  interaction limit of excitonic ordering.  At stoichiometry, the
  model reduces in the strong coupling limit to a pseudo-spin model
  with approximate U(4) symmetry.  Excitonic order appears in the
  pseudo-spin model as in-plane pseudo-magnetism.  The U(4) symmetry
  unifies all possible singlet and triplet order parameters describing
  such states.  Super-exchange, Hunds-rule coupling, and other
  perturbations act as anisotropies splitting the U(4) manifold,
  ultimately stabilizing a paramagnetic triplet state.  The tendency to
  ferromagnetism with doping (observed experimentally in the
  hexaborides) is explained as a spin-flop transition to a different
  orientation of the U(4) order parameter.  The physical mechanism
  favoring such a reorientation is the enhanced coherence (and hence
  lower kinetic energy) of the doped electrons in a ferromagnetic
  background relative to the paramagnet.  A discussion of the physical 
  meaning of various excitonic states and their experimental
  consequences is also provided.
\end{abstract}

\vspace{0.15cm}

\section{Introduction}
\label{sec:intro}

The unexpected discovery of high-$T_c$ itinerant ferromagnetism in
doped hexaborides\cite{hexaborides}\ has re-ignited interest in the
problem of excitonic ordering near the semiconductor--metal
transition.\cite{Keldysh,HalperinRice}\ Excitonically ordered states
are characterized by an off-diagonal order parameter describing
pairing between conduction electrons and valence holes.  Early
theoretical work by Volkov {\sl et. al.}\cite{Volkov}\ anticipated the
emergence of ferromagnetism on doping such an excitonic state.  These
authors considered the limit of nearly nested overlapping conduction
and valence bands with {\sl weak} repulsive electron-electron
interactions.  In this limit, the problem can be approximately cast
into a form nearly identical to BCS theory, and studied using the
techniques of mean-field theory.  Although this work (and some
subsequent recent studies\cite{Anisimov}) suffers from the important
physical mistake of neglecting the instability to phase separation,
ferromagnetism remains nevertheless a generic feature in a corrected
treatment.\cite{BalentsVarma,Barzykin}\ 

While the appearance of ferromagnetism in the weak-coupling limit is
encouraging, it is far from a conclusive and complete theoretical
explanation for the experiments.  First, Coulomb interactions in the
hexaborides are not particularly weak, and most likely are comparable
to the Fermi energy and band overlap.  Second, the above explanation
appears to hinge on the first-order nature of the excitonic to normal
(E-N) transition in the BCS limit.  While this feature, mathematically
analogous to the first-order transition to the normal state due to
pair-breaking by an external Zeeman field in a
superconductor,\cite{Fulde,LO}\ is present in the nested mean-field
limit, there do not appear to be any general theoretical grounds
mandating this behavior more generally.  Moreover, the universality of
the experimental results, now observed in a large number of different
compounds (Ca$_{1-x}$La$_x$B$_6$, BaB$_6$, Ca$_{1-x}$Ce$_x$B$_6$,
SrB$_6$,\ldots)\cite{private}, argues for
the robustness of the phenomenon.

\begin{figure}[h]
\begin{center}
\includegraphics[width=2.0in]{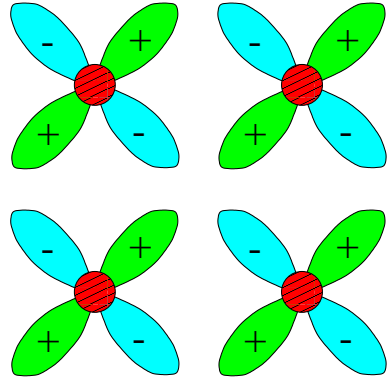}
\end{center}
\vspace{0.2in}
\caption{Imaginative illustration of a model for which the
  tight-binding description employed phenomenologically here
  directly applies. Red circles and blue+green crosses represent $s$
  and $d_{xy}$ orbitals, respectively.}
\label{fig:orbs}
\end{figure}

To determine whether excitonic ferromagnetism is indeed more general
than its weak-coupling theoretical basis, we consider here the
completely opposite {\sl strong-coupling} regime.  This is not
expected to be directly applicable to the hexaborides, as these
materials are most likely best described by an intermediate-coupling
model.  Nevertheless, many useful insights are gained from this
complementary limit.  As usual, the principle assumption of the
strong-coupling limit is the dominance of potential over kinetic
energy.  This is achieved concretely using a tight-binding model
(see Eqs.~\ref{tb1}-\ref{nnn}, in Sec.~\ref{sec:model}), in 
which the conduction and valence bands of the conventional continuum
theories are replaced by localized $a$ and $b$ orbitals, respectively.
The analog of band gap in the continuum model is the level splitting
$E_G = E_a-E_b > 0$.  The order parameter characterizing excitonic
ordering is then a matrix in spin space:
\begin{equation}
  \Delta_{\alpha\beta} = a_\alpha^\dagger
  b_\beta^{\vphantom\dagger},
  \label{deldef}
\end{equation}
where $a_\alpha^\dagger$ creates an electron with spin
$\alpha=\uparrow,\downarrow$ in the $a$ orbital, and
$b_\beta^{\vphantom\dagger}$ annihilates an electron with spin $\beta$
in the $b$ orbital.  Excitonically ordered states thus have some
partial occupation of the nominally excited $a$ states, as a result of
Coulombic repulsion.  In general, $\Delta_{\alpha\beta}$ is a proper
order parameter (i.e. one which characterizes a spontaneously broken
symmetry) if the $a$ and $b$ orbitals have different symmetries.  In
this paper, we consider a ``minimal model'' with this property,
comprised of one $a$ and one $b$ orbital per unit cell -- see
Fig.~\ref{fig:orbs}.  This mimics the situation in the hexaborides, for
which the conduction and valence states also transform as different
representations of the cubic point group\cite{Monnier}.  Because of
complications arising from orbital degeneracy, however, the
appropriate representations for the hexaborides are three-dimensional
rather than scalar.  We defer the possible complications arising from
these additional degrees of freedom to a future investigation.

As for the more familiar Hubbard model (see, e.g.
Ref.~\cite{Fradkin}), the problem simplifies somewhat in the
strong-coupling limit.  Considering first the undoped system
(half-filled = two electrons per unit cell), we obtain a novel quantum
pseudo-spin model (Eqs.~\ref{tJ1}-\ref{Ising},
Sec.~\ref{sec:undoped}).  Within this model, the excitonic insulator
(EI) appears as an intermediate state separating not a metal and a
semiconductor but a {\sl Mott insulator} and a semiconductor (or band
insulator).  In some respects, the behavior is argued to be quite
similar to that of a quantum spin-$1/2$ XXZ antiferromagnet in a
magnetic field, with excitonic ordering analogous to XY
antiferromagnetism.  The ``spins'' of the model, however, can take on
{\sl five} distinct states per site: one singlet state with both
electrons in the lower-energy $b$ orbital, and four different spin
states with one $a$ and one $b$ electron.  This is in contrast to the
two states of a single spin-$1/2$ particle.

In the strong coupling limit, this large Hilbert space is ``unified''
by several approximate symmetries valid at different energy scales.
At the largest energy scales this is an enormous SU(4) group,
corresponding to arbitrary complex rotations of the four components of
$\Delta_{\alpha\beta}$.  The approximate SU(4) symmetry fully unifies
all possible excitonic states, including singlet, triplet, and
singlet-triplet coexistences.  These are described by the general
decomposition 
\begin{equation}
  \Delta = {1 \over 2} \left( \Delta_s {\cal I} + \vec\Delta_t \cdot
    \vec\sigma^* \right),
  \label{ddecomp}
\end{equation}
where $\Delta_s$, $\vec\Delta_t$ are the singlet and triplet order
parameters, and ${\cal I}$ and $\vec\sigma$ are the $2\times 2$ unit
and Pauli matrices in spin space, respectively.  A system with
approximate SU(4) invariance contains the germ of ferromagnetism,
since several possible excitonic states (those with non-zero ${\rm
  Re}\, \Delta_s \vec\Delta_t^*$ and/or ${\rm Im}\, \vec\Delta_t
\wedge \vec\Delta_t^*$) give rise to net exchange fields, and hence a
magnetic moment.  SU(4) symmetry implies that these states are low in
energy.  At intermediate energies the SU(4) symmetry reduces to an
SU(2)$\times$SU(2) invariance, which reflects separate spin rotations
of the $a$ and $b$ electrons.  The latter is a symmetry of the
conventional continuum models of EIs, and transforms the order
parameter in a ``chiral'' manner: $\Delta \rightarrow U_L^\dagger
\Delta U_R^{\vphantom\dagger}$, where $U_L$ and $U_R$ are SU(2)
matrices.  Finally, further weak interactions reduce this to a simple
SU(2)$\times Z_2$ symmetry at the (very) lowest energies.

These symmetry considerations underly the simple physical mechanism
for ferromagnetism elucidated here\cite{so5}.  The dominant tendency
imposed by Coulomb interactions is to excitonic ordering.  With
approximate SU(4) symmetry, however, the ``orientation'' (form of
$\Delta_{\alpha\beta}$) of the order parameter is nearly free and
fixed only by weak ``anisotropy'' terms.  In the undoped material,
these anisotropies favor a simple paramagnetic triplet state.  Doping
introduces additional exchange energy contributions that modify the
anisotropy, causing $\Delta_{\alpha\beta}$ to ``flop'' into a
different orientation with a ferromagnetic moment.  In the present
model, the excitonic order in the ferromagnet is of non-collinear
triplet type, in which
\begin{equation}
  \Delta_s = 0, \qquad \vec\Delta_t \wedge \vec\Delta_t^* \neq 0.
\end{equation}
As shown in Sec.~\ref{sec:discussion}, in addition to ferromagnetic
magnetization, this state has additional spatially-varying local
static moments and spin currents transverse to the axis of net
magnetization.  The transition to this state from the paramagnet is
generally first order, and therefore coincides with a jump in the
electronic density.  Since experiments are performed at fixed charge
density (dictated by the concentration of dopant ions), the
intermediate ``forbidden'' range of dopings can be accommodated only by
phase separation.  With long-range Coulomb interactions included,
macroscopic phase separation is impossible, and charge domain
formation is expected, as pointed out already in
Refs.~\cite{BalentsVarma,Barzykin}.  

The {\sl detailed} demonstration of this behavior with doping is
non-trivial.  As for many other strongly correlated systems, the
problem of doping is much more difficult than that of the
stoichiometric Mott insulator.  Indeed, as the EI state lies
intermediate between band and Mott insulators, doping the EI is a sort
of interpolation between doping a conventional band insulator and
doping an antiferromagnetic insulator.  The latter problem is of
course at the crux of the physics of high-temperature
superconductivity, so that perhaps the experimental and theoretical
insights gained in the hexaborides will be helpful more generally.  At
any rate, doping the EI can be shown by very simple arguments to favor
ferromagnetism in strong coupling.  Essentially, the physics of this
behavior is similar to the ``Nagaoka effect''\cite{Nagaoka}\ in a doped
antiferromagnet -- ferromagnetic alignment of the excitonic order
parameters allows for more coherent propagation of the doped
electrons, and hence a lowering of their kinetic energy.  This
mechanism is actually {\sl stronger} in the EI than in the
antiferromagnet, because of the global {\sl coherence} of the
excitonic condensate, and the near degeneracy (due to approximate
SU(4) symmetry) of ferromagnetic and paramagnetic states.
\begin{figure}[h]
\begin{center}
\includegraphics[width=3.25in]{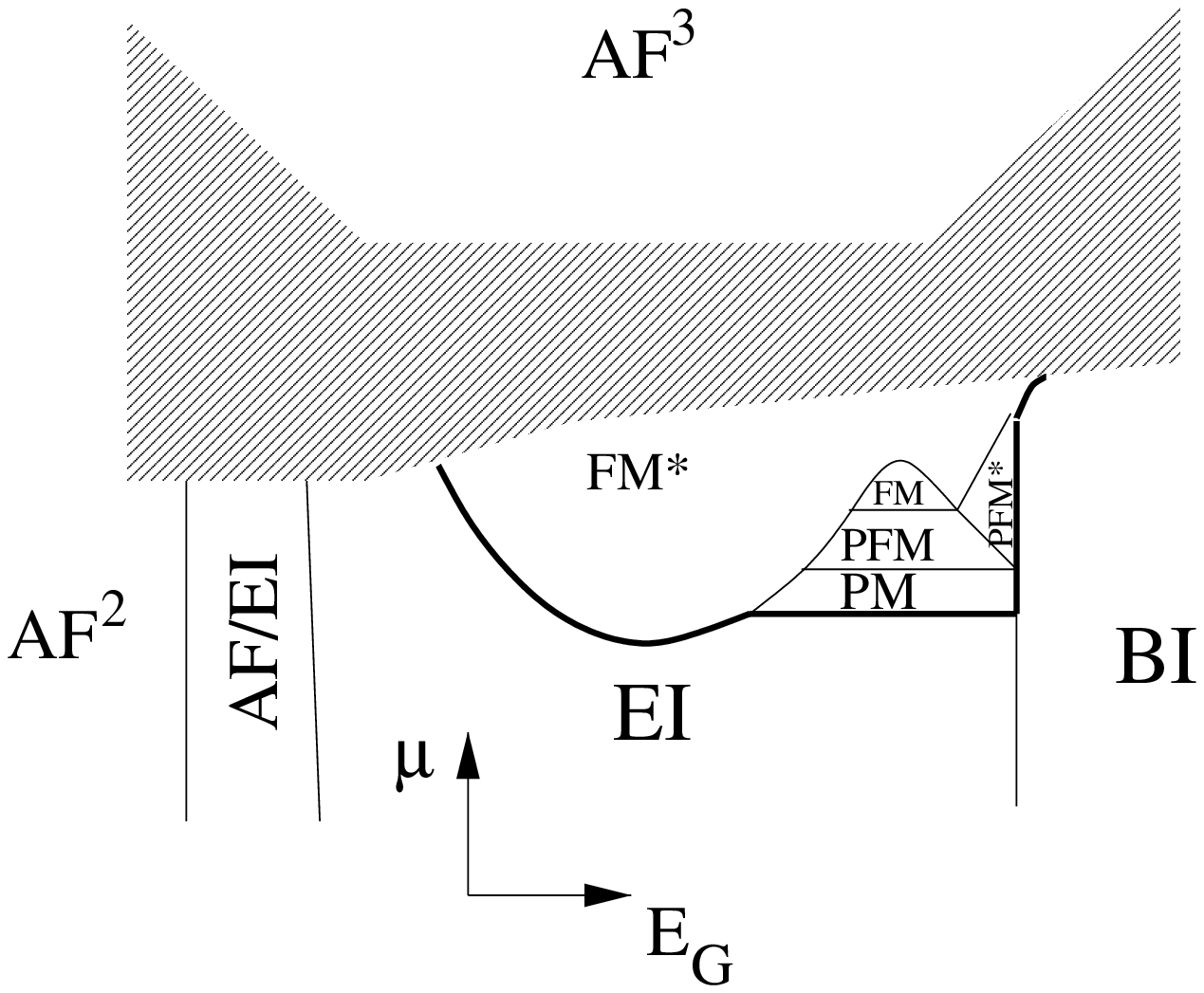}
\end{center}
\caption{Partial phase diagram of the strong-coupling model as a
  function of $E_{\rm\scriptscriptstyle G}$ (half the bare splitting
  between $a$ and $b$ orbitals), and $\mu$, the chemical potential.
  The shaded region is not analyzed in this paper.  Thick lines
  indicate the boundary between the undoped region below (with two
  electrons per unit cell) and the doped region above (with more than
  two electrons per unit cell).  The AF$^2$ and AF$^3$ phases are
  antiferromagnetic Mott insulators with two and three electrons per
  unit cell, respectively.  The BI state is the band insulator.
  Intermediate between the BI and AF$^2$ phases are the excitonic
  insulator (EI) and an insulator with coexisting excitonic and Ne\'el
  order (EI/AF).  The FM, FM$^*$, PFM, and PFM$^*$ phases are all
  ferromagnetic metals (see Table~\ref{tab:phases}\ for the
  differences between these states), while PM indicates a paramagnetic
  metallic phase.  All the metallic states above exhibit excitonic
  order.}
\label{partialfig}
\end{figure}
To provide a concrete demonstration of these ideas, the strong
coupling zero temperature phase diagram of the model is calculated in
this paper using a ``free Fermi gas'' approximation.  This
approximation captures the most important {\sl single quasiparticle}
physics of electronic propagation in an excitonically-ordered
background, but neglects interactions between these quasiparticles.
For simplicity, we also assume a {\sl fixed amplitude}, ${\rm Tr}\,
\Delta^\dagger \Delta^{\vphantom\dagger} = \Delta_0^2/2$, of the
excitonic order parameter.  The latter assumption is valid for weak
doping, $x \ll 1$, in which the {\sl orientation} of the ordering is
of paramount importance.  Putting together the results of this
calculation and the stoichiometric behavior, we arrive at the partial
phase diagram in Fig.~\ref{partialfig}.  This is in agreement with the
general expectations stated above.  It should be stressed, however,
that this analysis of doping is far from exhaustive.  More detailed
investigations of both the weak and strong coupling limits are
currently underway\cite{Veillette}\

The remainder of the paper is structured as follows.  In
Sec.~\ref{sec:model}, we present a detailed exposition of the
(simplest) tight-binding model capable of describing
excitonically-ordered states, and consider the limit of infinite
interaction strength.  The bulk of the paper is contained in
Sec.~\ref{sec:undoped}, where the model is analyzed for large but
finite interactions, focusing on the stoichiometric situation with two
valence electrons per unit cell.  For this electron density the model
is insulating, but can sustain excitonic and other types of ordering.
The properties of the model with doping are discussed in
Sec.~\ref{sec:doping}.  We conclude in Sec.~\ref{sec:discussion}\ with
a clarifying discussion delineating the physical properties of various
possible excitonic insulators, and the relation of the results of this 
paper to the hexaborides.

\section{Tight-Binding Model}

\label{sec:model}

\subsection{``On-site'' terms}

We consider a minimal model capable of exhibiting excitonic order,
which contains two orbitals per unit cell, so as to give rise to two
bands in a non-interacting limit (the actual situation in the
hexaborides is more complex, with orbital degeneracy leading to
multiple electron and hole pockets).  A strong-coupling limit is
obtained by first considering only {\sl local} interactions within a
unit cell:
\begin{eqnarray}
  H_0 & = & \sum_i E_{\scriptscriptstyle G} \left( a_i^\dagger
    a_i^{\vphantom\dagger} - b_i^\dagger b_i^{\vphantom\dagger}\right) - \mu 
  \left(a_i^\dagger a_i^{\vphantom\dagger} + b_i^\dagger
    b_i^{\vphantom\dagger}\right) 
  \nonumber \\
  & + & \!\!\!\! U\left( a_{i\uparrow}^\dagger a_{i\uparrow}^{\vphantom\dagger}
    a_{i\downarrow}^\dagger a_{i\downarrow}^{\vphantom\dagger} \!+\!
    b_{i\uparrow}^\dagger b_{i\uparrow}^{\vphantom\dagger}
    b_{i\downarrow}^\dagger b_{i\downarrow}^{\vphantom\dagger}\right) + V
  a_i^\dagger a_i^{\vphantom\dagger} b_i^\dagger
  b_i^{\vphantom\dagger}, 
  \label{tb1}
\end{eqnarray}
where $a,b$ are electron annihilation operators for the ``conduction'' 
and ``valence'' states, respectively, obeying
$\{a_{i\alpha}^{\vphantom\dagger},a_{j\beta}^\dagger\} =
\{b_{i\alpha}^{\vphantom\dagger},b_{j\beta}^\dagger\} =
\delta_{ij}\delta_{\alpha\beta}$.  Here and throughout the paper, we
use Latin indices $i,j,\ldots$ to denote the lattice site, and Greek
indices $\alpha,\beta,\ldots=\uparrow,\downarrow$ to denote the spin
state.  Labels will be suppressed and implicit wherever clarity
allows.   The parameters $E_{\scriptscriptstyle
  G}$,$\mu$,$U$,$V$ describe the ``band gap'' (orbital energy
difference), chemical potential, on-site ``Hubbard'' repulsion, and
nearest neighbor repulsion, respectively, within the unit cell.  

A crucial feature of $H_0$ is the absence of direct hopping between
the $a$ and $b$ orbitals within the unit cell.  For excitonic ordering
to be well-defined, it is necessary at a minimum that the $a$ and $b$
states be distinguished by a discrete symmetry operation, e.g. parity.
When this is the case, direct hopping between these orbitals is
prohibited.  It may be helpful to imagine an artificial situation in
which the $a$ and $b$ orbitals represent $s$ and $d_{xy}$ orbitals on
a single site of a square lattice (see Fig.~\ref{fig:orbs}).

In this situation, $a$ and $b$ orbitals are orthogonal both on the
same site and on nearest neighbor sites.  An overlap {\sl is}
possible, though for next-nearest neighbor pairs, i.e. on a diagonal.
In general, an exchange interaction {\sl is} allowed by symmetry, and
takes the form
\begin{equation}
  H_1 = -J_H \sum_i \vec{S}_{ia}\cdot \vec{S}_{ib},
  \label{tb2}
\end{equation}
where $\vec{S}_{ia} = {1 \over 2}a_i^\dagger
\vec{\sigma}a_i^{\vphantom\dagger}$, $\vec{S}_{ib} = {1 \over 2}
b_i^\dagger \vec{\sigma} b_i^{\vphantom\dagger}$.  Here and in the
following, the Pauli matrices $\vec{\sigma}$ act in the spin space.
On physical grounds, a ferromagnetic exchange ($J_{H}>0$) is most
appropriate due to Hund's rule effects for orthogonal orbitals.  For
pedagogical purposes, we may wish to consider instead the opposite
antiferromagnetic sign for this exchange.  From the discussion in
Sec.~1, it is clear that an essential ingredient for excitonic
ferromagnetism is the near-degeneracy of singlet and triplet states.
To build this into the strong coupling model thus requires small
$J_{H}$.  For the majority of the paper, therefore, we will neglect
$J_{H}$ or treat it as a small perturbation.

\subsection{Infinite interaction limit}

\begin{figure}[h]
\begin{center}
\includegraphics[width=2.75in]{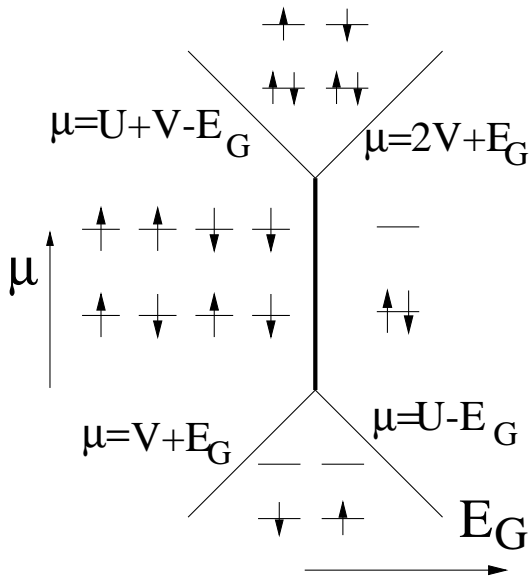}
\end{center}
\vspace{0.2in}
\caption{Strong coupling (ultra-local) phase diagram in the $E_{\rm
    \scriptscriptstyle G}$--$\mu$ plane, neglecting exchange and all
  inter-cell hopping and interactions.  Regions with zero and four
  electrons per unit cell are not shown.  In each phase, the
  lowest-energy states are pictured, with $a$ orbital above and $b$
  orbital below.}
\label{fig:ultralocal}
\end{figure}

The analysis of the strong-coupling limit begins by first considering
the on-site Hamiltonian, $H_{\rm site} = H_0+H_1$, in the absence of
electron hopping between adjacent unit cells.  This may be thought of
as the analog of the $U = \infty$ analysis of an ordinary Hubbard
model.  In this case, the occupation of each orbital is a good quantum
number, and the states can be straightforwardly enumerated.  Assuming
$E_{\rm \scriptscriptstyle G}>V>0$, and at first also $J_{ab}=0$, one
obtains the phase diagram shown in Fig.~\ref{fig:ultralocal}.
For the present study, we are particularly interested in densities
near two electrons per unit cell.  Note that the doping behavior (i.e.
on increasing $\mu$) in this regime depends crucially on the relative
strength of $E_{\rm \scriptscriptstyle G}$ and $U$.  In particular,
for $2E_{\rm \scriptscriptstyle G} >U-V$, the preferred charge $Q=2e$
state is one with both electrons in the lower orbital, corresponding
to the band insulator.  For $U-V>2E_{\rm \scriptscriptstyle G}$, by
contrast, the two-electron ground state has one electron in each 
orbital, and hence a net spin on each site.  This is the ultra-strong
coupling (i.e. local) version of a Mott insulator.  Note that {\sl
  neither} of these two states exhibits {\sl excitonic} order.  This
can be seen by directly computing $\langle a^\dagger
b^{\vphantom\dagger}\rangle = 0$ in either state.  In fact, the
operators $a^\dagger b^{\vphantom\dagger}$ and $b^\dagger a$ act to
transform the two phases into one another, i.e. move an electron from
the lower to upper orbital or vice versa.  
\subsection{Hopping terms}

To investigate further, we must introduce hopping 
between adjacent cells.  We will principally consider the simplest
such term,
\begin{equation}
  H^\prime = \sum_{\langle ij\rangle} t \left( a_i^\dagger
    a_j^{\vphantom\dagger} + b_i^\dagger b_j^{\vphantom\dagger} + {\rm 
      h.c.}\right),
  \label{tb3}
\end{equation}
where $\langle ij\rangle$ indicates that the sum is over nearest
neighbor pairs of sites.  Different hopping integrals $t_a$ and $t_b$
between the two orbitals can also be easily included, but do not
change the results significantly, so we will keep $t_a=t_b=t$ for
simplicity (see however, the discussion of particle-hole symmetry in
Sec.~\ref{sec:discussion}\ surrounding Eqs.~\ref{ph1}-\ref{ph2}).  In
general, there are also hopping processes connecting $a$ and $b$
orbitals.  Due to the symmetry of the orbitals in Fig.~1, these occur
only for next-nearest neighbors,
\begin{equation}
  H^{\prime\prime} = \sum_{\langle\langle ij\rangle\rangle} t_{ab}
  {\rm sign}[(x_i-x_j)(y_i-y_j)] \left(
    a_i^\dagger b_j^{\vphantom\dagger} + {\rm h.c.}\right),
  \label{nnn}
\end{equation}
where the double angular brackets denote a sum over next nearest
neighbors $i$ and $j$.  Note that the hopping matrix elements in
Eq.~\ref{nnn}\ are real and vary in sign.  The sign variations reflect
the symmetry differences (under rotations) between the $s$ and $d$
orbitals.  The reality of the coefficients is a matter of convention,
which we fix by choosing the orbital wavefunctions to be real.  We
will assume, as appropriate in this example, that $t_{ab} \ll t$, so
that $H^{\prime\prime}$ is small, and can therefore be treated
perturbatively.

It is sometimes an important perturbation, because it reduces the
symmetry of the Hamiltonian.  In particular, all of the terms in
$H_0+H_1+H'$ conserve the number of $a$ and $b$ particles separately.
Neglecting the $a$--$b$ hopping, therefore, the model has $SU(2)\times
U(1)\times U(1)$ continuous symmetries, corresponding to conservation
of spin, and $a$ and $b$ charges.  The perturbation $H^{\prime\prime}$
reduces the continuous symmetries of the model down to $SU(2)\times
U(1)$ corresponding to spin and total charge, which are required by
the physics of the system.  Eq.~\ref{nnn}\ actually still respects a
number of discrete symmetries, such as $b \rightarrow -b$
simultaneously with a $\pi/2$ rotation.  These symmetries, which in
fact comprise the point-group operations of the square lattice, can be
viewed as a residual discrete subgroup of the original $U(1)$ present
in the absence of $H^{\prime\prime}$.  We will see in the next section
that this gives rise to an Ising symmetry under which the excitonic
order parameters transform.

\section{Effective Theory for the Undoped System}

\label{sec:undoped}

In the central region of Fig.~2, e.g. for $(U+V)/2 < \mu < (U+3V)/2$,
all sites are doubly occupied in the strong coupling limit.
Nevertheless, for $2E_G \leq U-V$, the low energy states are highly
degenerate.  Well to the left of the thick vertical line, each $a$ and
$b$ orbital is singly occupied, so that there are effectively two
spin-$1/2$ degrees of freedom in each unit cell.  In the infinite
coupling limit these are completely free, but they will interact due
to virtual hopping processes when $H^\prime$ (and $H^{\prime\prime}$)
is included.  Far to the right of the vertical line, the unique low
energy state consists of a doubly occupied $b$ orbital in each unit
cell, and hopping is unimportant.  As the vertical line is approached
from either side, virtual hopping processes can induce interactions
involving all {\sl five} low energy states.  

\subsection{Bosonic $t$--$J$ model}

In this subsection, we develop an effective model for the interesting
region near the vertical line.  In this region, it is necessary and
sufficient to truncate the Hilbert space to just the five low-energy
states in each unit cell (although higher energy states must be kept
in virtual processes).  
The physics is amusingly
similar to a sort of generalized bosonic $t$--$J$ model.  On the
left-hand side of the thick vertical line, each unit cell is occupied by two
spins.  At second order in $H^\prime$, these interact via effective
exchange interactions,
\begin{equation}
  H_{\rm eff}^{s}  =  \sum_{\langle ij \rangle} J\left(
    \vec{S}_{ia}\cdot\vec{S}_{ja} +
    \vec{S}_{ib}\cdot\vec{S}_{jb}\right) - \sum_i J_H
  \vec{S}_{ia}\cdot\vec{S}_{ib},
  \label{tJ1}
\end{equation}
where $J = t^2/(V+2E_{\rm \scriptscriptstyle G})$.  This exchange
constant may be obtained by computing perturbatively the energy
difference between singlet and triplet states on a bond to second
order in the hoppings, and neglecting the deviation from the vertical
line (i.e. setting $U=2E_{\rm \scriptscriptstyle G}+V$) in the
denominators.  The latter approximation is valid provided
$|U-V-2E_{\rm \scriptscriptstyle G}| \ll V$.  Well to the left of the
vertical line (in particular when $U-V-2E_{\rm \scriptscriptstyle G}
\gg t$), no doubly-occupied $b$ states are present, and
Eq.~\ref{tJ1}\ is a complete model.  It describes two
ferromagnetically bulk coupled Heisenberg spin-$1/2$
antiferromagnets.  On a hypercubic lattice (square or cubic in two or
three dimensions, respectively), one expects long-range
antiferromagnetic order of spins on the same orbital sublattice, with
$a$ and $b$ spins aligned parallel at each site.

As the vertical line is approached, the energy cost of a
doubly-occupied $b$ orbital is reduced towards zero, and they must be
introduced into the lattice.  Unit cells with both electrons in the
$b$ orbital act as ``holes'', having no associated local moment.
Unlike the usual $t$-$J$ model holes, they are, however, bosonic and
neutral (relative to the magnetic state, they represent the removal of 
an $a$ electron and replacement with a $b$ electron).  Hole hopping
occurs at second order in $t$, :
\begin{eqnarray}
  H_{\rm eff}^{h} & = & - \mu_h \sum_i
  h_i^\dagger h_i^{\vphantom\dagger} +  t_h \sum_{\langle ij \rangle}  \left(
    h_i^\dagger h_j^{\vphantom\dagger} + h_j^\dagger
    h_i^{\vphantom\dagger}\right) {\cal P}_{ij} \nonumber \\
  & & + \sum_{\langle ij \rangle} V_{hh} \,h_i^\dagger 
  h_i^{\vphantom\dagger} h_j^\dagger h_j^{\vphantom\dagger}
  \label{tJ2}
\end{eqnarray}
where $\mu_h = 2E_{\rm \scriptscriptstyle G} - U + V + t^2/(2V) -
t^2/[2(2E_{\rm \scriptscriptstyle G}+V)]$ is the hole
``chemical potential'', $t_h = t^2/(2V)$, $V_{hh} =  t^2/V - {1 \over
  2}t^2/(2E_{\rm \scriptscriptstyle G}+V)$, 
and ${\cal P}_{ij} = ({3 \over 2}+ 
2\vec{S}_{ia}\cdot\vec{S}_{ja})({3 \over 2} +
2\vec{S}_{ib}\cdot\vec{S}_{jb})$ is the operator which interchanges the
spin states at sites $i$ and $j$.  Like in a conventional doped
anti-ferromagnet, the presence of the ${\cal P}_{ij}$ operator in the
hopping term leads to difficulties of hole motion in an
antiferromagnetic spin background.  Naive successive hopping of a
single hole in an antiferromagnetic state results in a generalization
of the well-known ``string'' of misaligned spins in its wake.  

Introducing the $a-b$ hopping term (Eq.~\ref{nnn}) affects the system
in several ways.  There are renormalizations of the coupling constants
in Eq.~\ref{tJ2}\ and Eq.~\ref{tJ1}, of order $t_{ab}^2/V$,
$t_{ab}^2/(V+4E_G)$. Since, by assumption, $t_{ab} \ll t$, these are
negligible.  New exchange couplings are also generated between
next-nearest-neighbor $a$ and $b$ spins, which were not previously
present.  Because they are small, unfrustrating, and break no
additional symmetries, these are also negligible.  The most important
effect is to introduce a term which violates $h$--particle
conservation:
\begin{equation}
  H_{\rm eff}^{nnn} = \sum_{\langle\langle ij\rangle\rangle} y \bigg[
  h_i h_j \big(\vec{O}^{t\dagger}_{a;ij}\cdot
  \vec{O}^{t\dagger}_{b;ij} + O^{s\dagger}_{a;ij}
  O^{s\dagger}_{b;ij}\big) + {\rm h.c.} \bigg].
  \label{fugacity}
\end{equation}
Here $\vec{O}^{t\dagger}_{a/b;ij}$ creates a triplet of spin one
states of $a/b$ particles on the pair of sites $ij$, $O^{s\dagger}_{a/b;ij}$
creates a singlet of $a/b$ particles on this pair, and the
``fugacity'' $y= 2t_{ab}^2/V$.  Note that although Eq.~\ref{fugacity}\ 
violates conservation of the number of ``holes'', it creates and
annihilates them only in pairs.  There thus remains a conserved Ising
charge or parity ($= \sum_i h_i^\dagger h_i^{\vphantom\dagger} {\rm
  (mod 2)}$), signifying whether the number of holes is even or odd.
This parity can be traced back to fact that the two orbitals on each
site transform differently under spatial reflections. 

\subsection{Pseudo-spin description}

To understand the behavior of this model, we now introduce a useful
reformulation.  Formally, the five possible states on each site can be
viewed as different quantized values of a generalized pseudo-spin, and
the above terms then take the form of nearest-neighbor interactions
between these spins.  In particular, we define five states per site
via $|1\rangle = a_\uparrow^\dagger b_\uparrow^\dagger |v\rangle,
|2\rangle = a_\uparrow^\dagger b_\downarrow^\dagger |v\rangle,
|3\rangle = a_\downarrow^\dagger b_\uparrow^\dagger |v\rangle,
|4\rangle = a_\downarrow^\dagger b_\downarrow^\dagger |v\rangle,
|5\rangle = b_\uparrow^\dagger b_\downarrow^\dagger |v\rangle$.  The
Hamiltonian can be rewritten in terms of $5\times 5$ spin matrices
${\cal T}^{\mu\nu}$, where $\langle \mu'|{\cal T}^{\mu\nu}|\nu'\rangle
= \delta^{\mu\mu'}\delta^{\nu\nu'}$. Neglecting for the moment the
hole non-conserving terms in Eq.~\ref{fugacity}, $H_{\rm eff}^h =
H_{\rm eff}^{ps} + {\rm const.}$, where
\begin{equation}
  H_{\rm eff}^{ps}\! =\! \sum_{\langle ij\rangle} {{\cal J}_\perp
    \over 2}
  \! \sum_{\mu=1}^4 \! \left({\cal T}^{\mu 5}_i {\cal T}^{5\mu}_j  \!
    + \!  i \leftrightarrow j\right) \! +\! {\cal J}_z {\cal T}_i^z
  {\cal T}_j^z \! - \!
  {\cal H} \sum_i {\cal T}_i^z,
  \label{pseudospin}
\end{equation}
and ${\cal T}^z_i = (\sum_{\mu=1}^4 {\cal
  T}_i^{\mu\mu}- {\cal T}^{55}_i )/2$.  The generalized exchange
constants ${\cal J}_\perp = 2t_h$, ${\cal J}_z=V_{hh}$, and Zeeman
field ${\cal H}=dV_{hh}/2 - \mu_h$. 

This form of the Hamiltonian exposes a strong similarity to the
spin-$1/2$ XXZ model in a Zeeman field.  In particular, the ``boson
hopping'' ${\cal J}_\perp$ is analogous to an
antiferromagnetic in-plane exchange ($S_i^+ S_j^-$ terms ), spin-boson 
interaction ${\cal J}^z$ to an antiferromagnetic Ising exchange, and
${\cal H}$ to a $z$-axis field.  For ${\cal J}_\perp \gg {\cal J}_z$ and
${\cal H}$
not too large, one expects the analog of canted XY antiferromagnetism, 
while for ${\cal J}_z \gg {\cal J}_\perp$, one expects instead
$z$-axis Ising antiferromagnetism up to a threshold value of $|{\cal H}|$.  
For large fields, $|{\cal H}| \gg {\cal J}_\perp,
{\cal J}_z$, one expects ultimately fully polarized states, which 
correspond to the Mott and band insulators for ${\cal H}>0$ and ${\cal 
  H}<0$,
respectively.  

Surprisingly, $H_{\rm eff}^{ps}$ displays an enormous SU(4) invariance 
under $T^{5\mu} \rightarrow \sum_{\nu=1}^4 U_{\mu\nu}T^{5\nu}$,
$T^{\mu 5} \rightarrow \sum_{\nu=1}^4 U^*_{\mu\nu} T^{\nu 5}$, where
$U$ is an SU(4) matrix.  SU(4) symmetry is expected to be
a good approximation over a range of energies, because in the 
physical limit $V \ll U \sim E_{\rm \scriptscriptstyle G}$, $J_H \ll J \ll
{\cal J}_\perp, {\cal J}_z, {\cal H}$.  Thus we will take the approach of
first solving the SU(4) invariant model, and considering successively
the exchanges $J$ and $J_H$, which reduce the symmetry of $H_{\rm
  eff}$ to SU(2)$\times$SU(2) (independent {\sl physical} spin
rotations of the $a$ and $b$ moments) and SU(2)$\times$U(1),
respectively.  

Lastly, we consider the effects of the hole-pair creation and
annihilation terms in Eq.~\ref{fugacity}, which can also be
transcribed into the pseudo-spin language.  One finds $H_{\rm
  eff}^{ps} \rightarrow H_{\rm eff}^{ps} + H_{\rm eff}^I$, where
\begin{eqnarray}
  H_{\rm eff}^I & = & \sum_{\langle\langle ij\rangle\rangle} {\cal
    J}_I \bigg[ {\cal T}^{25}_i {\cal T}^{25}_j + {\cal T}^{35}_i
    {\cal T}^{35}_j \nonumber \\
    & & - {\cal T}^{15}_i {\cal T}^{45}_j - {\cal
      T}^{45}_i {\cal T}^{15}_j + 
    {\cal T}^{52}_i {\cal T}^{52}_j + {\cal T}^{53}_i
    {\cal T}^{53}_j \nonumber \\
    & & - {\cal T}^{51}_i {\cal T}^{54}_j - {\cal
      T}^{54}_i {\cal T}^{51}_j \bigg].
    \label{Ising}
\end{eqnarray}
The coupling ${\cal J}_I \propto y$.
While it is perhaps not completely transparent in this notation (a
better notation for this term will be introduced in next subsection -- 
see Eq.~\ref{Ising2}),
the effect of $H_{\rm eff}^I$ is to further break the
SU(2)$\times$U(1) symmetry down to SU(2)$\times$Z$_2$.  The Z$_2$
invariance is the remnant of the physical parity symmetry discussed in 
the previous subsection.

\subsection{Mean-field theory and undoped phase diagram}

\begin{figure}[h]
  \setlength{\unitlength}{1.0in}
  \begin{picture}(4.0,3.0)(0,0)
    \put(0.4,0.1){\includegraphics[width=3.0in]{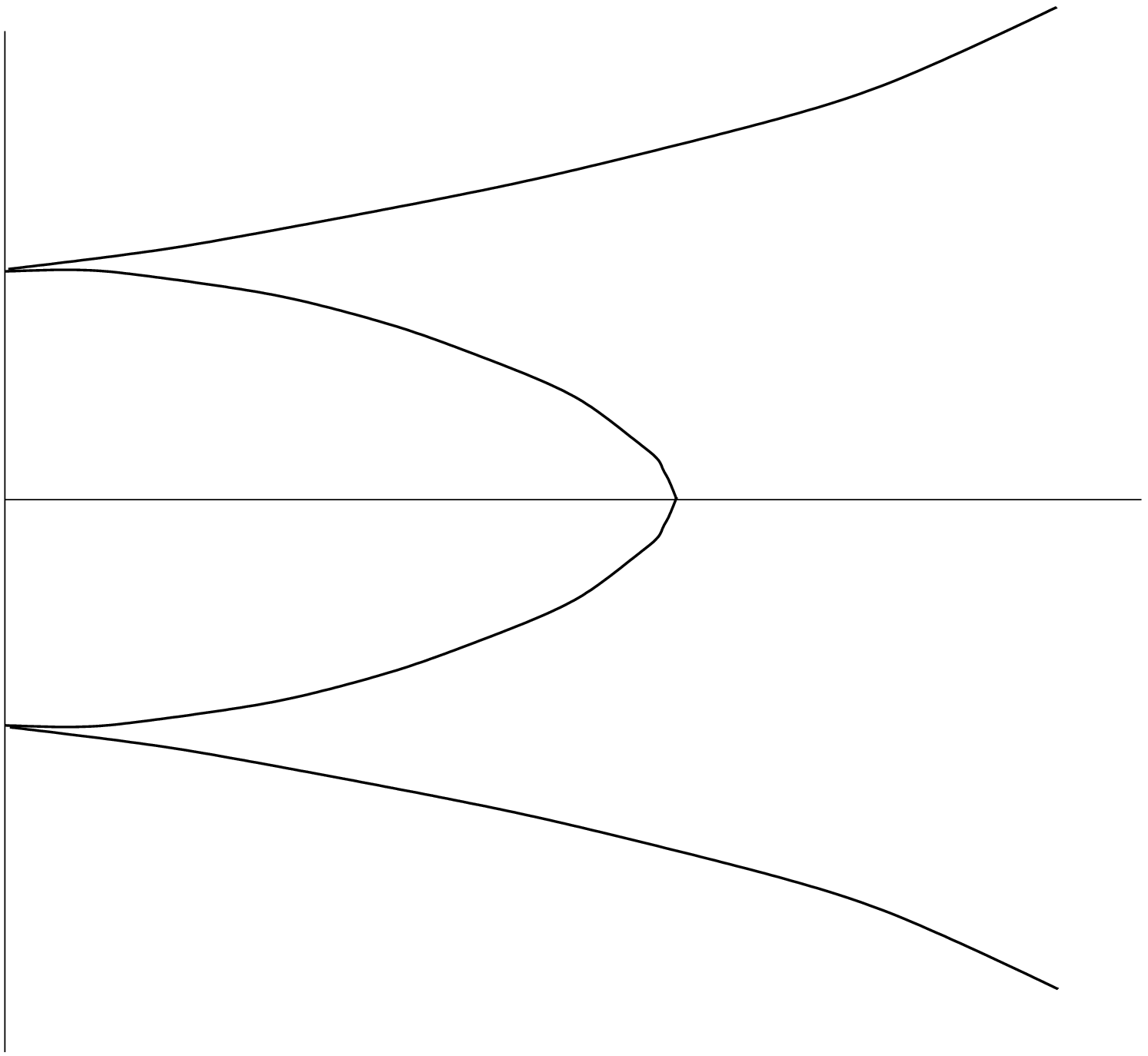}}
    \put(0.0,2.8){\large ${\cal H}/{\cal J}_z$}
    \put(2.8,1.7){\large ${\cal J}_\perp/{\cal J}_z$}
    \put(2.6,2.1){\Large \bf EI}
    \put(1.0,1.7){\Large \bf MPS}
    \put(1.0,2.6){\Large \bf MI $\uparrow$}
    \put(1.0,0.5){\Large \bf BI $\downarrow$}
  \end{picture}
\caption{Mean-field phase diagram of the fully SU(4)-invariant pseudo-spin
  model.  A large pseudo-Zeeman field (which scales linearly with the
  orbital splitting $E_G$) stabilizes either the Mott insulating (MI)
  or band insulating (BI) state, depending upon its sign.  For
  ``in-plane'' anisotropy (${\cal J}_\perp >{\cal J}_z$), the
  intervening phase is an excitonic insulator (EI).  In the opposite
  limit (``Ising'' anisotropy) it consists of a micro-phase-separated
  state with a checkerboard pattern of alternating band and Mott
  insulating configurations at the lattice scale.  In obtaining the
  pseudo-spin model from the strong-coupling limit of
  Eqs.~\ref{tb1}-\ref{tb3}, one finds in-plane anisotropy, and the
  intermediate state is expected to be excitonically ordered.  }
\label{XXZfig}
\end{figure}

We expect that a simple Weiss mean field theory (MFT) gives
qualitatively correct results for the stoichiometric phase diagram, as 
it does for the ordinary XXZ+Zeeman model.\cite{largeNnote}\
Neglecting $H_{\rm eff}^s$, the MFT consists in replacing     
\begin{equation}
  T_i^{\mu 5} T_j^{5 \mu} \rightarrow \langle T_i^{\mu5 }\rangle
  T_j^{5 \mu} + T_i^{\mu 5} 
  \langle T_j^{5 \mu}\rangle - \langle T_i^{\mu 5}\rangle \langle
  T_j^{5 \mu}\rangle,
\end{equation}
for each bond $i,j$ on the lattice, and similarly for the $T^z_i
T^z_j$ interaction.  With this replacement, the Hamiltonian decouples
on different lattice sites, and the problem reduces to solving
self-consistently the appropriate single-site problems.  As an
antiferromagnetic solution is expected, this amounts to equations for
the ($8$ component) transverse staggered magnetization, defined by
$\langle T_i^{\mu 5}\rangle = (-1)^i [n_\perp^{2\mu-1} + i
n_\perp^{2\mu}]$, and the uniform and staggered $z$-axis magnetizations,
defined by $\langle T_i^z\rangle = m_z + (-1)^i n_z$.  Because of SU(4)
symmetry, all orientations of $n_\perp^k$ are degenerate, and it is
sufficient to assume $n_\perp^k \equiv n_\perp \delta^{k 1}$.  In
this subspace, the equations of MFT become {\sl identical} to those of
the conventional spin-$1/2$ XXZ antiferromagnet in a Zeeman field.
These equations were solved in Ref.~\cite{XXZ}.  The resulting
phase diagram is shown in Fig.~\ref{XXZfig}.
%

Since ${\cal J}_\perp > {\cal J}_z$, we expect transverse pseudo-spin
polarization, $\langle {\cal T}^{\mu 5} \rangle \neq 0$, provided
$|{\cal H}| < {\cal H}_c = d{\cal J}_\perp$.  Remarkably, the
transverse components of the pseudo-spin operator are exactly the
excitonic order parameters.  In particular, straightforward algebra
shows ${\cal T}^{\mu 5}  = (-\Delta_{\uparrow\downarrow},
\Delta_{\uparrow\uparrow}, -\Delta_{\downarrow\downarrow},
\Delta_{\downarrow\uparrow})$.  Thus for ${\cal J}_\perp > {\cal
  J}_z$, MFT predicts an excitonic insulator.

We now turn to the evolution of the ground state in this regime on
introducing the symmetry-breaking terms in $H_{\rm eff}^s$.  In their
absence, the excitonic order parameter can ``point'' in any direction
which is equivalent under the broken SU(4) symmetry.  Within MFT, this
amounts to complete freedom to choose the four complex components of
$\Delta_{\alpha\beta}$, subject to the constraint ${\rm Tr} \,
\Delta^\dagger \Delta = {1 \over 4}(1- {\cal H}^2/{\cal H}_c^2) \equiv
\Delta_0^2/2$.  In term of singlet and triplet components defined by
Eq.~\ref{ddecomp}, this constraint simply implies $|\Delta_s|^2 +
\vec{\Delta}_t^*\cdot\vec{\Delta}_t = \Delta_0^2$.  The perturbations
in $H_{\rm eff}^s$ can be viewed as ``anisotropies'' favoring
sub-manifolds within this space.  

To clarify the nature of the anisotropy terms,
it is helpful to work with the mean-field wavefunction,
$|\Psi_0\rangle = \prod_i {\cal E}_i^\dagger | BI \rangle$, where
$|BI\rangle = \prod_i b_{i\uparrow}^\dagger b_{i\downarrow}|v\rangle$
is the non-interacting band-insulating state, and
\begin{equation}
  {\cal E}^\dagger_i = c\left( 1 + (-1)^i|c|^{-2}\sum_{\alpha\beta}
    \Delta_{\alpha\beta}^* a_{i\alpha}^\dagger b_{i\beta}^{\vphantom\dagger} 
    \right)
    \label{creationop}
\end{equation}
is a local ``exciton creation operator''.  Here $|c|^2 = (1-{\cal
  H}/{\cal H}_c)/2$.  It is now straightforward to evaluate the
expectation value of $H_{\rm eff}^s$ in the mean-field ground state.
Up to a constant for fixed ${\rm Tr}\, \Delta^\dagger
\Delta^{\vphantom\dagger}$, on a hyper-cubic lattice one finds the bulk
energy density
\begin{equation}
  \epsilon_b \equiv L^{-d}\langle H_{\rm eff}^s\rangle = 
    2\tilde{J}\, {\rm Tr} \left(\Delta^\dagger
        \Delta^{\vphantom\dagger}\right)^2 + \tilde{J}_H  \left| {\rm Tr}
        \, \Delta \right|^2,
      \label{anisotropy}
\end{equation}
where $\tilde{J} = da^{-d} J/2|c|^4$ and $\tilde{J}_H =
a^{-d}J_H/2|c|^2$.  The above terms are essentially completely
determined by the SU(2)$\times$SU(2) and SU(2)$\times$U(1) symmetries.
To proceed, we employ two identities derivable from 
Eq.~\ref{ddecomp}:
\begin{eqnarray}
  {\rm Tr}\, \left(\Delta^\dagger \Delta^{\vphantom\dagger} \right)^2
  & = & {1 \over 8} \left(\Delta_s^* \Delta_s^{\vphantom *} +
    \vec\Delta_t^* \cdot \vec\Delta_t^{\vphantom *} \right)^2
  \nonumber \\
  & + & {1 \over 8} \left| \Delta_s^* \vec\Delta_t^{\vphantom *} +
    \Delta_s^{\vphantom *} \vec\Delta_t^* - i \vec\Delta_t^{\vphantom *} \wedge
    \vec\Delta_t^* \right|^2, \label{id1} \\
  {\rm Tr}\, \Delta & = & \Delta_s \label{id2} .
\end{eqnarray}
By assumption, $J \gg J_H$, so that the first term in
Eq.~\ref{anisotropy}\ creates the dominant splitting of the SU(4)
ground-state degeneracy.  The low-energy sub-manifold thus consists of
the order parameters which minimize ${\rm Tr} (\Delta^\dagger
\Delta^{\vphantom\dagger})^2$.  Eq.~\ref{id1}\ then implies
$\Delta_s^* \vec\Delta_t^{\vphantom *} + \Delta_s^{\vphantom *}
\vec\Delta_t^* - i \vec\Delta_t^{\vphantom *} \wedge \vec\Delta_t^*
=0$ (note that the first term in Eq.~\ref{id1}\ is constant and equal
to $\Delta_0^2/8$).  The physical content of this condition is made
clear by calculating the mean spin polarization on the $a$ site using
the mean-field wavefunction in Eq.~\ref{creationop}:
\begin{equation}
  \vec{s}_a = \langle \vec{S}_a \rangle = {1 \over {4|c|^2}} \left( i \vec\Delta_t^{\vphantom
      *} \wedge  \vec\Delta_t^* + \Delta_s^*
    \vec\Delta_t^{\vphantom *} + 
    \Delta_s^{\vphantom *} \vec\Delta_t^* \right).
  \label{meansa}
\end{equation}
Thus the influence of the exchange coupling $J$ is to favor states
with $\vec{s}_a = 0$.  

This condition still allows a fairly large range of states, the
simplest of which are pure singlet ($|\Delta_s| = \Delta_0$,
$\vec\Delta_t = 0$) and pure collinear triplet ($\Delta_s =0$,
$\vec\Delta_t \neq 0$, $\vec\Delta_t^{\vphantom
      *} \wedge  \vec\Delta_t^* = 0$) orderings.  The additional
effect of the Hunds-rule ferromagnetic coupling $J_H$ is to introduce
a small extra ``mass'' for the singlet order parameter, favoring a
pure triplet state.

The {\sl phase} of the triplet order parameter is determined
by the ``Ising anisotropy'' terms in Eq.~\ref{Ising}.  To see this, we 
rewrite ${\cal T}^{\mu 5}$ and ${\cal T}^{5\mu}$ directly in terms of
$\Delta$.  One finds
\begin{equation}
   H_{\rm eff}^I = \sum_{\langle\langle ij\rangle\rangle} {\cal
    J}_I \, {\rm Tr}\, \left( \Delta_i^{\vphantom\dagger}
    \Delta_j^{\vphantom\dagger} + \Delta^\dagger_j \Delta^\dagger_i
  \right). \label{Ising2}
\end{equation}
Note that Eq.~\ref{Ising2}\ explicitly breaks the U(1) symmetry of
phase rotations of $\Delta$, down to the Ising invariance $\Delta
\rightarrow - \Delta$.  If $H_{\rm eff}^I$ is considered a weak
perturbation, it can be treated by simply evaluating its expectation
value in the mean-field ground state (Eq.~\ref{creationop}), giving
$\langle \Delta_i \rangle = \langle \Delta_j \rangle = \Delta$, since
$i$ and $j$ are next-nearest neighbors.  Using ${\rm Tr} \, \Delta^2 =
(\Delta_s^2 + \vec{\Delta}_t\cdot \vec{\Delta}_t)/2$, one finds (since
${\cal J}^I >0$) that Eq.~\ref{Ising2}\ favors an {\sl imaginary}
triplet order parameter $\vec\Delta_t = - \vec\Delta_t^*$.  This is
{\sl different} from the weak-coupling treatment of
Ref.~\cite{BalentsVarma}, in which a {\sl real} triplet order
parameter was found to be preferred.

Unlike in superconductivity, the phase of the excitonic order
parameter has physical significance, as discussed by Halperin and
Rice\cite{HalperinRice}.  In particular, it is straightforward to show 
that a {\sl real} $\vec\Delta_t$ order parameter corresponds to a
non-zero average spin density within the unit cell of the crystal, while for
$\vec\Delta_t$ imaginary, the spin density is zero but there are
instead non-zero spin {\sl currents}.  The imaginary triplet state
obtained here is therefore a sort of spin ``flux phase'' with non-zero 
spin currents.  See Sec.~\ref{sec:discussion}\ for a more in-depth
discussion.  

Apart from this difference, the strong-coupling results of this
section are in very close agreement with the weak-coupling results of
Ref.~\cite{BalentsVarma}.  Indeed, not too much significance should be
attached to the difference in phase of the order parameters, as indeed
the models are in any case not completely identical.  In fact, the
detailed correspondence of results up to this point strongly argues
for a continuous smooth interpolation (``adiabatic continuity'') of
most physical properties of such systems as the overall interaction
strength is increased from small to large values.

Finally, we comment on the modifications to the SU(4)-invariant phase
diagram in the presence of the symmetry-breaking terms in
Eqs.~\ref{tJ1},~\ref{fugacity}.  As argued above, these favor an
imaginary triplet state when ${\cal H}=0$.  Inside the Mott insulator,
these terms stabilize an antiferromagnetically ordered magnetic state.
On approaching the Mott insulator boundary, therefore, we expect the
emergence of magnetic ordering.  This implies the existence of at
least one additional phase boundary separating the triplet EI (which
has no non-zero spin density) from a magnetically ordered EI with
non-zero average spin density, somewhere inside the region in which
the EI phase occurs in the SU(4)-invariant model.

\section{Doping}

\label{sec:doping}

In this section, we consider the behavior as a low density of electrons is
added to the system.  In the strong-coupling limit, this reduces to an
effective $t$-$J$--like model, in which the Hilbert space is
restricted to states in which all sites (unit cells) are either doubly
(corresponding to the excitonic pseudo-spins modeled above) or triply
occupied, the latter containing one $a$ and two $b$ electrons.  The
system is then governed by an effective Hamiltonian $H_{\rm dope} =
H_{\rm eff}^s + H_{\rm eff}^{ps} + \tilde{\cal P}H'\tilde{\cal P}$,
where $\tilde{\cal P}$ projects onto this restricted Hilbert space.

As many years of work on high-$T_c$
superconductivity has taught us, the problem of doping a correlated
(Mott) insulator, particularly with spin (and here pseudospin)
ordering, is extremely complex and difficult.  Here, we will adopt the 
absolute simplest approach extending the above MFT to the low electron 
density limit.  We assume, as suggested by the weak-coupling analysis, 
that the essential ingredient for excitonic ferromagnetism is the
approximate enhanced (in this case SU(4)) symmetry of the effective
Hamiltonian.  In considering the doped state, then, it is crucial to
determine in what way the added electrons affect the splitting of the
degenerate SU(4) ground-state manifold.

\subsection{Variational treatment for a single electron}

In the strong-coupling limit, the majority of the energy of an added
electron is kinetic, since $t \gg {\cal J}, {\cal J_\perp} \sim
t^2/V$, etc..  Just as in the simpler but much studied $t$--$J$ model
for the cuprates,  coherent motion of an added electron, however, is
greatly hindered by (pseudo)-spin ordering of the insulating
background.  Moreover, coherent motion is possible to a varying degree
depending upon the precise nature of the background.  We first
consider this effect for a single added electron using the variational
method.  A natural variational ansatz is 
\begin{equation}
  |\Psi_1\rangle = \sum_{i\alpha} \psi_{i\alpha}^{\vphantom\dagger}
  a_{i\alpha}^\dagger \prod_{j\neq i} {\cal E}_j^\dagger |BI\rangle,
\end{equation}
where {\sl both} the doped electron's wavefunction $\psi_{i\alpha}$
and the excitonic order parameter $\Delta_{\alpha\beta}$ (implicit in
${\cal E}_j^\dagger$) are considered as variational parameters.  For
fixed ${\rm Tr}\, \Delta^\dagger \Delta$, the energy depends only upon 
$J$, $J_H$, and $t$.  In particular, one
finds 
\begin{equation}
  \epsilon_v = L^{-d} \langle \Psi_1|H_{\rm dope}|\Psi_1\rangle =
  \epsilon_b(1-2 d (a_0/L)^d )  +  L^{-d}\epsilon_e,
  \label{evar}
\end{equation}
where $a_0$ is the lattice spacing,
\begin{equation}
  \epsilon_e = t \sum_{\langle ij\rangle}
  \psi_{i\alpha}^* \hat{T}_{\alpha\beta} \psi_{j\beta}^{\vphantom\dagger},
  \label{hopping}
\end{equation}
and the matrix $\hat{T}_{\alpha\beta} =
|c|^2 \delta_{\alpha\beta} + |c|^{-2} 
\left( \Delta^{\vphantom\dagger}\Delta^\dagger
\right)_{\alpha\beta}$.  Physically, we identify the first term in
Eq.~\ref{evar}\ as the bulk energy density, reduced by the presence of 
a single doped electron (occupying the volume fraction $(a/L)^d$).
In the second term, the quantity $\epsilon_e$ is then readily interpreted as
the energy of the added electron.  Eq.~\ref{hopping}\ is then a
hopping Hamiltonian for this electron.  In a polarized
excitonic background, this hopping is in general non-diagonal in spin.
In terms of singlet and triplet components, 
\begin{equation}
  \hat{T} = {(1+|c|^2) \over 2} + \vec{s}_a\cdot \vec{\sigma}^* ,
  \label{Tmatrix}
\end{equation}
where  $\vec{s}_a$, the mean spin polarization on the $a$ site, is
given by Eq.~\ref{meansa}.
Minimizing Eq.~\ref{hopping}\ in the space of normalized wavefunctions 
$\psi_{i\alpha}$ gives the tight-binding Schr\"odinger equation,
\begin{equation}
  t \sum_{\langle ji\rangle} \sum_{\beta}  \hat{T}_{\alpha\beta}
  \psi_{j\beta} = \epsilon_e \psi_{i\alpha},
\end{equation}
where the angular brackets indicate a sum over the nearest neighbors
$j$ of site $i$.  The single-particle eigenstates of this equation are
plane waves with spins polarized parallel and antiparallel to
$\vec{s}_a$, with eigenvalues
\begin{equation}
  \epsilon_{e\pm}({\bf k}) = 2 t\left[{{1+|c|^2} \over 2} \pm
    |\vec{s}_a|\right] \sum_{i=1}^d \left[\cos k_i a_0 \right],
  \label{vardisp}
\end{equation} 
where $a_0$ is the lattice spacing.  The location of the
minimum-energy electronic excitations depends crucially on the
magnitude of $\vec{s}_a$, and hence ${\cal H}$.  When ${\cal H}>{\cal
  H}_c/3$, electrons with spin parallel and antiparallel to
$\vec{s}_a$ have minimum energy at different points in momentum space.
Such large values of ${\cal H}$ correspond to strongly overlapping
bands, close to the boundary between the Mott and excitonic
insulators.  For simplicity, we will specialize to the case when
$|\vec{s}_a| < (1+|c|^2)/2$, which occurs for ${\cal H} < {\cal
  H}_c/3$.  In this case, the minimal energy single-particle energy
excitations for both spin orientations have momentum ${\bf k} =
(\pi,\cdots,\pi)$.  Furthermore, the optimal spin orientation is
parallel to $\vec{s}_a$.  Such an electron takes advantage of the
``Zeeman'' energy due to the exchange field (proportional to
$\vec{s}_a$) generated by the ``core'' spins (i.e. the spins of the
two electrons per unit cell present in the insulator).

In the undoped system, however, $\vec{s}_a = 0$, due to the anisotropy
in Eq.~\ref{anisotropy}. We therefore expect that the optimal order
parameter in the doped system is determined by a competition between
these two terms.  With some algebra, it is straightforward to verify
that, due to the Hunds-rule term $J_H$, the complex pure triplet state
(i.e. with $\vec\Delta_t\wedge\vec\Delta_t^* \neq 0$ but $\Delta_s=0$)
is always more energetically favorable than a singlet-triplet
coexistence (with ${\rm Re}\, \Delta_s^* \vec\Delta_t \neq
0$).\cite{differentnote}\ Without loss of generality, it is thus
convenient to choose a spin quantization axis, letting
\begin{equation}
  \vec{\Delta}_t = \Delta_0(\cos\theta {\bf\hat{x}} +
  i\sin\theta{\bf\hat{y}}),
  \label{nonunitary}
\end{equation}
One then finds $\vec{s}_a = -(\Delta_0^2/2|c|^2) \sin (2\theta)
{\bf\hat{z}}$.  In any such state, $\vec{s}_b = \vec{s}_a$, so that
the core spins also contribute to the ferromagnetic moment.  

\subsection{Free Fermi gas approximation}

It remains to determine the optimal angle $\theta$.  To proceed, we
need to extend Eq.~\ref{evar}\ to a small but non-zero {\sl density}
of doped electrons.  At low densities, it seems natural to neglect
interactions between doped electrons, and use the simplest possible
{\sl free Fermi gas} estimate for the electronic dopant energy.  In
particular, we approximate the energy of the system as the sum of two
contributions: a ``bulk'' contribution from the undoped unit cells
containing two electrons and a spatially uniform order parameter
$\Delta_{\alpha\beta}$, and a ``dopant'' contribution, approximated by
the energy of a free Fermi gas of electrons with dispersion given by
Eq.~\ref{vardisp}.  For concreteness, the detailed formulae are
presented in the following for three spatial dimensions ($d=3$).  At
low densities, only single-particle states near ${\bf k} = \boldsymbol{\pi}
= (\pi,\pi,\pi)$ are occupied, so it is convenient to expand around
this point, ${\bf k} = \boldsymbol{\pi} + {\bf q}$, yielding the dispersion
\begin{equation}
  \epsilon_{e\pm}({\bf q}) = - 2t\left[ {(1+|c|^2) \over 2} \pm |\vec{s}_a|
  \right] \left[ 3 - {{q^2 a_0^2} \over 2} \right] - \tilde\mu.
\end{equation}
Here we have re-instated a (shifted) chemical potential $\tilde\mu$
to control the density of doped electrons.  It is both convenient and
physically helpful to work at fixed chemical potential rather than
fixed charge density, as this allows naturally for the possibility of
phase separation.  As is perhaps not surprising based on the results
of weak-coupling analysis\cite{BalentsVarma,Barzykin}, we will see that
phase separation does indeed occur in a physically interesting
parameter range of the model (at least within this approximation).

\begin{figure}[h]
  \setlength{\unitlength}{1.0in}
  \begin{picture}(4.0,3.0)(0,0)
    \put(0.3,0.5){\includegraphics[width=3.0in]{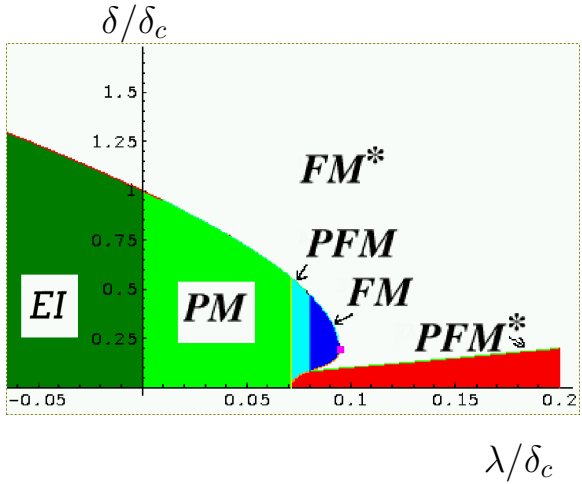}}
    \put(0.8,2.5){\Large $\delta/\delta_c$}
    \put(2.8,0.2){\Large $\lambda/\delta_c$}
  \end{picture}
\caption{Free-Fermi gas approximation to the zero-temperature phase
  diagram.  The abscissa
  $\lambda/\delta_c$ is the non-dimensionalized chemical potential,
  and the ordinate $\delta/\delta_c$ indicates the strength of
  excitonic ordering relative to super-exchange interactions (see
  Eqs.~\ref{ef}-\ref{gammadef}\ for the precise definitions). The
  properties of the various phases shown are listed in
  Table~\ref{tab:phases}.  For larger $\lambda$ (not shown), both
  excitonic ordering and ferromagnetism disappear, owing to reduction
  of the amplitude of $\Delta$.}
\label{dlfig}
\end{figure}

Because we are interested in the energy density only insofar as to
determine the angle $\theta$, we neglect in the following all terms
independent of $\theta$.  Inserting Eq.~\ref{nonunitary}\ into
Eq.~\ref{anisotropy}\ gives the bulk energy 
\begin{equation}
\epsilon_b = 
[3 J \Delta_0^4/(8a_0^3 |c|^4) ]\sin^2 2\theta + {\rm const.}
\end{equation}
(in three dimensions).  This must be added to the ground state energy
of the free Fermi gas of doped electrons.  Simple but tedious
algebraic calculations lead to the final expression for the total energy
density of the system:
\begin{equation}
  \epsilon_f = \overline{\epsilon} \delta^2 \left[ g^2 -
    \sqrt{\delta \over \delta_c} {\cal E}\left(g,
      \gamma\right) \right],  
  \label{ef}
\end{equation}
where 
\begin{eqnarray}
  \overline\epsilon & = & {{3J(1+|c|^2)^2} \over 8a^3}, \\
  \delta & = & {\Delta_0^2 \over {|c|^2(1+|c|^2)}}, \\
  \delta_c & = & \left( {5 \pi^2 (1+|c|^2) \over {16\sqrt{6}}} {J \over t}
  \right)^2, \\
  g & = & \sin 2\theta  \\ 
  \lambda & = & 1 + {\tilde\mu \over {3t(1+|c|^2)}}, \\
  \gamma & = & \lambda/\delta. \label{gammadef}
\end{eqnarray}
The function ${\cal E}(g,\gamma)$ is straightforwardly related to the
energy density of the three-dimensional free electron gas in a Zeeman
field.  In general it depends not only on $g$ and $\gamma$, but also
on $\delta$.  For simplicity, we will assume $|\delta g| \ll 1$, which 
holds near to the excitonic insulator--band insulator boundary, and is
satisfied more generally in the interesting region of the phase
diagram  (where $\delta$ is $O(\delta_c)$, since $\delta_c \ll 1$ in
the strong coupling limit $J/t\ll 1$ -- see Fig.~\ref{dlfig}).  In
this case, ${\cal E}(g,\gamma)$ becomes independent of $\delta$.  Its
functional form  is
\begin{equation}
  {\cal E}(g,\gamma) = \sum_{z=\pm 1} (\gamma + zg)^{5/2}
  \Theta(\gamma+zg),
  \label{cale}
\end{equation}
where $\Theta(\gamma)$ is the Heavyside step function.
Eqs.~\ref{ef}--\ref{cale}\ give the energy density of the system as a
function of chemical potential $\tilde\mu$ (through $\gamma$) and order
parameter angle $\theta$ (through $g$).  The optimal excitonic angle
$\theta$ is determined by 
minimizing $\epsilon(\tilde\mu,\theta)$ at fixed $\tilde\mu$.  If
$\tilde\mu$ and $\theta$ are known, the density of doped electrons $x$ 
and itinerant magnetization density $m_{it}$ are then given by the
free-fermion 
results: 
\begin{eqnarray}
  x & = & {\sqrt{2} \over {6\pi^2 a^3}}\sum_{z=\pm 1} (\gamma + z g)^{3/2}
    \Theta(\gamma + z g), \\
  m_{it} & = & {1 \over {6\sqrt{2} \pi^2 a^3}} \sum_{z=\pm 1} z (\gamma +
  z g)^{3/2} \Theta(\gamma + z g).
  \label{mag}
\end{eqnarray}
Note that the system is doped (i.e. $x\neq 0$) whenever $\gamma+|g| >
0$.  One should keep in mind also that the full magnetization density
$m= m_{it}+m_{core}$ includes a contribution $m_{core} =
-(\Delta_0^2/|c|^2a^3) \sin (2\theta)$ from the core
spins.

Eqs.~\ref{ef}-\ref{mag}\ completely determine the state of the system
at zero-temperature as a function of $\lambda$ and $\delta$.  The
mathematical problem of minimizing $\epsilon_f$ is algebraically quite
tedious, and significant care must be taken to avoid spurious local
minima and saddle points.  The results of a careful study are shown in
Fig.~\ref{dlfig}.  All the phases shown are excitonically ordered, but
differ in doping $x$, excitonic angle $\theta$, and magnetization $m$.
The properties of each are summarized in Table~\ref{tab:phases}.  In the strong-coupling limit, we expect (see Sec.~\ref{sec:doping})
$\delta/\delta_c \gg 1$, in which case there is a direct {\sl
  first-order} transition from a paramagnetic excitonic insulator (EI)
to a fully-polarized ferromagnetic metal (FMFP$^*$).  
The phase boundaries in Fig.~\ref{dlfig}\ variously indicate first
(discontinuous) and second(discontinuous) order transitions.  All the
vertical phase boundaries denote continuous transitions, while most of
the transitions on curved phase boundaries are discontinuous.  The
exceptions are the PPFM$^*$--FPFM$^*$ boundary (which is everywhere
second order) and the lower-portion of the FPFM--FPFM$^*$ transition
line, which is continuous below the tricritical point indicated in the
figure.

\begin{table}[h]
  \begin{tabular}{ccccc}
   {\bf phase} & {\bf doped?} & {\bf mag.} & {\bf pol.} &
   {\bf angle} \\ 
    \hline EI & no & para & n/a & $\theta=0$ \\
    PM & yes & para & n/a & $\theta=0$ \\
    PFM & yes & ferro & partial & $0<\theta<\pi/4$ \\
    FM & yes & ferro & full & $0<\theta<\pi/4$ \\
    PFM$^*$ & yes & ferro & partial & $\theta=\pi/4$ \\    
    FM$^*$ & yes & ferro & full & $\theta = \pi/4$ \\
  \end{tabular}
  \caption{Phases of the doped EI in the free-Fermi gas
    approximation.  The five columns list the abbreviation, presence
    or absence of doping, magnetic order, degree of polarization, and
    excitonic angle ($\theta$), respectively, for the six phases.}
  \label{tab:phases}
\end{table}

Which portion of this phase diagram is most physically significant?
In the strong-coupling limit, $\delta_c \ll 1$, and it therefore seems 
reasonable to suppose $\delta/\delta_c \gg 1$, so that the system
undergoes a simple and direct first order transition from the undoped
and paramagnetic EI to the fully-rotated half-metallic ferromagnet,
FPFM*.  Coincident with this transition is a jump in the electronic
charge density $x$, from zero in the insulator to a non-zero value in
the metal.  

\section{Discussion}
\label{sec:discussion}

\subsection{Symmetries and properties of excitonic insulators}

The model introduced in Sec.~\ref{sec:model}\ contains many possible
excitonically ordered states in various regions of its phase diagram.
In the undoped case,  we have argued that a simple paramagnetic
collinear triplet ordering is most likely, while a state with
$\vec\Delta_t\wedge \vec\Delta_t^* \neq 0$ obtains for electron
densities slightly greater than two per unit cell.  Nevertheless, if,
as supposed, SU(4) symmetry is a good approximation, then many other
possible states must necessarily be nearly as low in energy.  In the
hope that the truth may ultimately be decided by experimental
measurements, it seems useful to delineate the physical
characteristics of each of these phases.  

With the exception of the non-collinearly ordered states, the analysis
of the next few paragraphs is identical (though in somewhat different
notation) to that of Halperin and Rice\cite{HalperinRice}.  First, let
us consider the existence of a time-averaged magnetic moment.  In the
tight-binding formulation, the electron field operator is expanded in
terms of Wannier orbitals,
\begin{equation}
  \psi_\alpha({\bf r}) = \sum_i \left[ \phi_a({\bf r} - {\bf R}_i)
    a_{i\alpha} + \phi_b({\bf r}-{\bf R}_i) b_{i\alpha} \right],
  \label{decomp}
\end{equation}
where $\phi_{a/b}({\bf r})$ is the Wannier function for the $a/b$
orbital, and we neglect the other (unoccupied) states.  Consider next
the spin density operator.  We will assume for simplicity (though this
is not essential) that each Wannier function has support only within
one unit cell.  Eq.~\ref{decomp}\ then leads to a representation for
the spin density operator $\vec{S}$,
\begin{eqnarray}
  2\vec{S}({\bf r}) & = & \psi^\dagger \vec{\sigma}
  \psi^{\vphantom\dagger}, \\
  &  = & |\phi_a({\bf r})|^2 \langle a^\dagger \vec{\sigma}
    a^{\vphantom\dagger}\rangle + |\phi_b({\bf r})|^2 \langle
    b^\dagger \vec{\sigma}  b \rangle \nonumber \\
  &  + &\phi_a^*({\bf r}) \phi_b({\bf r}) \langle a^\dagger \vec{\sigma}
  b^{\vphantom\dagger} \rangle + \phi_b^*({\bf r}) \phi_a({\bf r})
  \langle b^\dagger \vec{\sigma} a^{\vphantom\dagger} \rangle. \label{spineqns}
\end{eqnarray}
To proceed, we choose both Wannier functions to be {\sl real}.  Then
{\sl for the undoped case}, the spin density can be rewritten in terms of
$\vec{s}_{a/b}$ and $\vec\Delta_t$:
\begin{equation}
  \vec{S}({\bf r}) = |\phi_a({\bf r})|^2 \vec{s}_a + |\phi_b({\bf
    r})|^2 \vec{s}_b + 2 \phi_a({\bf r})\phi_b({\bf r}) {\rm Re}\,
  \vec\Delta_t. 
\end{equation}
Recall further Eq.~\ref{meansa}\ and its analog for $\vec{s}_b$:
\begin{equation}
  \vec{s}_{a/b} = {1 \over {4|c|^2}}
  \left[i \vec\Delta_t^{\vphantom
      *} \wedge  \vec\Delta_t^* \pm \left(\Delta_s^* 
    \vec\Delta_t^{\vphantom *} + 
    \Delta_s^{\vphantom *} \vec\Delta_t^*\right) \right].
\label{sab}
\end{equation}
There are thus non-zero static local moments whenever ${\rm Re}\,
\Delta_s^* \vec{\Delta}_t$, ${\rm Im}\,
\vec\Delta_t\wedge\vec\Delta_t^*$, or ${\rm Re}\, \vec\Delta_t$ are
non-zero.  In the simplest such states, $\vec{\Delta}_t =
|\vec{\Delta}_t| \hat{e}$, where $\hat{e}$ is a real unit vector.  In
this case, there is a spatially-varying static moment within the unit
cell oriented along the $\hat{e}$ axis.  The {\sl net} moment
(integrated over the unit cell) is, however, zero, unless
$\vec\Delta_t \wedge \vec\Delta_t^* \neq 0$, in which case the real
and imaginary parts of $\vec\Delta_t$ are both non-zero and not
parallel.  In addition to the net ferromagnetic polarization along
${\rm Im}\, \vec\Delta_t \wedge \vec\Delta_t^*$, such states have a
non-collinear static spin density in the unit cell.  The net moment
along these other directions remains zero.  To see why such states
sustain a net polarization, consider the particular case given in
Eq.~\ref{nonunitary}, with excitonic angle $\theta$.  One can then use
Eq.~\ref{ddecomp}\ to rewrite the order parameter matrix as
\begin{equation}
  \Delta = {\Delta_0 \over 2}\left[(\cos\theta+\sin\theta) \sigma^+ +
    (\cos\theta-\sin\theta) \sigma^-\right].
  \label{rewrite}
\end{equation}
Inspection of the mean-field wavefunction, Eq.~\ref{creationop}\ and
Eq.~\ref{rewrite}\ immediately shows that the amplitude for up and
down spins are unequal,  so long as $\theta$ is not a multiple of
$\pi$.  

Some confusion may arise in the reader with regard to time-reversal
symmetry.  It appears surprising to have $\vec\Delta_t$ and
$i\vec\Delta_t\wedge \vec\Delta_t^*$, the latter containing a
cross-product, both contributing to $\vec{s}_{a/b}$.  In fact, both
terms transform like a spin under time-reversal.  This is simplest to
see in the path-integral representation of the quantum system, in
which the Fermion operators are replaced by time-dependent Grassman
fields $a^{\vphantom\dagger}_\alpha \rightarrow
a^{\vphantom\dagger}_\alpha(t)$, $a_\alpha^\dagger \rightarrow
\overline{a}_\alpha(t)$, and similarly for
$b^{\vphantom\dagger}_\alpha, b_\alpha^\dagger$.  The Grassman fields
then transform under time-reversal according to
\begin{eqnarray}
  a_\alpha(t) & \rightarrow & \sigma^y_{\alpha\beta}
  \overline{a}_\beta(-t), \\
  \overline{a}_\alpha(t) & \rightarrow & - \sigma^y_{\alpha\beta}
  a_\beta(-t), \\
  b_\alpha(t) & \rightarrow & \sigma^y_{\alpha\beta}
  \overline{b}_\beta(-t), \\
  \overline{b}_\alpha(t) & \rightarrow & - \sigma^y_{\alpha\beta}
  b_\beta(-t).
\end{eqnarray}
Note the important minus sign in the above transformation, which is
possible because $a_\alpha$, $\overline{a}_\alpha$ ($b_\alpha$,
$\overline{b}_\alpha$) are independent 
fields (not related by complex conjugation) in the path integral.
This reflects the anti-unitary nature of time-reversal symmetry.  At
any rate, Eqs.~\ref{deldef}-\ref{ddecomp}\ then imply that 
\begin{eqnarray}
  \Delta_s & \rightarrow & \Delta_s^*, \\
  \vec\Delta_t & \rightarrow & -\vec\Delta_t^*, 
\end{eqnarray}
under time-reversal.  The combination of complex conjugation and the
minus sign for $\vec\Delta_t$ imply that both terms in Eq.~\ref{sab}\
are odd under time-reversal.  Indeed, the necessary and sufficient
conditions for broken time-reversal symmetry is ${\rm Re}\,
\langle \vec\Delta_t\rangle \neq 0$ and/or ${\rm Im}\,
\langle\Delta_s\rangle \neq 0$.

A perhaps surprising consequence of Eqs.~\ref{spineqns}\ is that
apparently if ${\rm Re}\Delta_s^* \vec\Delta_t \neq 0$ but ${\rm Im}\,
\vec\Delta_t \wedge \vec\Delta_t^* =0$, there is no net magnetization.
In fact, this result applies only to the particular undoped model
consider here, and is a consequence of a special variety of
particle-hole symmetry (which we denote PH).  To make this explicit,
define a hole creation operator $\tilde{b}_\alpha^\dagger =
\sigma^y_{\alpha\beta}b_\beta$.  Then the electron number operator can
be rewritten as
\begin{equation}
  n = a^\dagger a^{\vphantom\dagger} + b^\dagger b^{\vphantom\dagger}
  = 2+ a^\dagger a^{\vphantom\dagger}-\tilde{b}^\dagger
  \tilde{b}^{\vphantom\dagger}. 
  \label{ph1}
\end{equation}
In the undoped system, the mean number of electrons per unit cell is
two, so that $\langle a^\dagger a^{\vphantom\dagger}-\tilde{b}^\dagger
\tilde{b}^{\vphantom\dagger}\rangle = 0$.  Thus precisely at this
density, and {\sl only} at this density, we may entertain the
possibility of symmetry under the transformation PH:
\begin{equation}
  a_\alpha \rightarrow_{PH} \tilde{b}_\alpha, \qquad \tilde{b}_\alpha
  \rightarrow_{PH} a_\alpha.
\end{equation}
In the new variables, the excitonic order parameter becomes $\Delta =
a_\alpha^\dagger \sigma^y_{\alpha\beta} \tilde{b}_\beta^\dagger$.
Thus $\Delta \rightarrow -\Delta^T$ (here the superscript $T$
indicates the matrix transpose) under PH.  Also useful is the operator
$a^\dagger a^{\vphantom\dagger} - b^\dagger b^{\vphantom\dagger} =
a^\dagger b^{\vphantom\dagger} + \tilde{b}^\dagger
\tilde{b}^{\vphantom\dagger} -2$ (proportional to ${\cal T}^z$ in the
$n=2$ subspace), which is invariant under PH.  Thus $H_{\rm eff}^{ps}$
(see Eq.~\ref{pseudospin}) is PH-invariant.  Similarly, it is
straightforward to show that under PH, the two spin operators are
exchanged: 
\begin{equation}
  \vec{S}_a \leftrightarrow_{PH} \vec{S}_b.
  \label{spinic}
\end{equation}
Thus $H_{\rm eff}^s$ is also PH-invariant, as is $H_{\rm eff}^I$, as
can be easily shown.  Thus the undoped Hamiltonian is invariant under
PH.  Considering the order parameters, we find that
\begin{eqnarray}
  \Delta_s & \rightarrow_{PH} & \Delta_s, \\
  \vec\Delta_t & \rightarrow_{PH} & - \vec\Delta_t.
  \label{ph2}
\end{eqnarray}
Thus the combination ${\rm Re}\,\Delta_s^* \vec\Delta_t$ is {\sl odd}
under PH, and hence cannot give rise to a total moment, since
$\vec{S}_{\rm TOT}$ is PH-invariant.  ${\rm Im}\,
\vec\Delta_t \wedge \vec\Delta_t^*$, however, is PH-invariant, and can 
hence couple directly to a ferromagnetic moment.

It should be stressed that PH {\sl is not a microscopically exact
  symmetry}, even in the stoichiometric situation.  It occurred in the
above analysis only because of the arbitrary choice of equal hopping
between $a$ and $b$ orbitals, $t_a=t_b = t$, in Eq.~\ref{tb3}.  In
general, one expects $t_a \neq t_b$, which leads to different
anti-ferromagnetic exchange constants between $a$ and $b$ spins in
$H_{\rm eff}^s$, Eq.~\ref{tJ1}.  Different exchange constants destroy
the invariance of the Hamiltonian under the interchange of $a$ and $b$
spins, Eq.~\ref{spinic}, which is the effect of PH.  It is
straightforward to show that, when this asymmetry is included in the
microscopic Hamiltonian, states with ${\rm Re}\Delta_s^* \vec\Delta_t
\neq 0$ are also ferromagnetic.  In addition, even in the model with
$t_a=t_b$, {\sl doping} breaks the PH symmetry, and gives rise to a
ferromagnetic moment in the  ${\rm Re}\Delta_s^* \vec\Delta_t
\neq 0$ state.

Considerations similar to those above Eq.~\ref{spineqns}\ apply to the
electronic charge density ($\rho$), current density ($\vec I$) , and
spin current density ($J^{\mu\nu}$) operators.  One finds
\begin{eqnarray}
  \rho({\bf r}) & = & -e|\phi_a({\bf r})|^2 n_a -e|\phi_b({\bf r})|^2
  n_b \nonumber \\
  & & -2e \phi_a({\bf r})\phi_b({\bf r}){\rm Re}\, \Delta_s, \nonumber 
  \\
  \vec{I}({\bf r}) & = & {e \over m} {\rm Im}\, \Delta_s \left[
    \phi_a({\bf r})
    \vec\nabla \phi_b ({\bf r}) - \phi_b({\bf r})
    \vec\nabla \phi_a ({\bf r})  \right], \nonumber \\
  J^{\mu\nu}({\bf r}) & = & \!\!\!\!\!{1 \over 2m} {\rm Im}\, \Delta_t^\mu
  \left[ \phi_a({\bf r})
    \partial_\nu \phi_b ({\bf r})\! -\! \phi_b({\bf r})
    \partial_\nu\phi_a ({\bf r})  \right].
  \label{currentsetc}
\end{eqnarray}
In the final equation above, $J^{\mu\nu}$ is the current density for
spin polarized along the $\mu$ axis propagating in the $\nu$
direction.  For completeness, the mean-field expressions for the
number of $a$ and $b$ particles are
\begin{eqnarray}
  n_a & = & \langle a^\dagger a^{\vphantom\dagger}\rangle = |c|^{-2}
  {\rm Tr}\, \Delta^\dagger \Delta^{\vphantom\dagger}, \\
  n_b & = & \langle b^\dagger b^{\vphantom\dagger}\rangle = 2-n_a.
\end{eqnarray}

From Eqs.~\ref{currentsetc}, we can read off the physical
interpretation of the various other types of ordering.  If ${\rm
  Im}\,\vec\Delta_t \neq 0$, there is a spontaneous {\sl spin current} 
in the unit cell.  This is necessarily the case for any state with
$\vec\Delta_t \wedge \vec\Delta_t^* \neq 0$, which, as discussed
above, also exhibits non-collinear static moments.  The simpler state
with $\vec\Delta_t = i |\Delta_t| \hat{e}$ has {\sl only} the spin
currents, and is the magnetic analog of a ``flux phase'' in modern
terminology.  Similarly, if the singlet order parameter has an
imaginary part ${\rm Im} \Delta_s \neq 0$, there are non-zero charge
currents within the unit cell.  This is exactly a flux phase.
Finally, a real singlet order parameter, $\Delta_s = \Delta_s^* \neq
0$, gives rise to a charge-density $\rho({\bf r})$ that breaks the
point group symmetry of the crystal, since $\phi_a({\bf r})
\phi_b({\bf r})$ is not a scalar.  

Another importance characteristic of the phases with triplet ordering
is a finite (transverse) uniform spin susceptibility.  This is a very
general consequence of broken spin-rotational invariance.  In the
simplest collinear triplet states, $\Delta_t = \Delta_0 e^{i\phi}
\hat{e}$, where $\hat{e}$ is a real vector.  The elementary
excitations of the symmetry-broken state can then be classified {\sl
  only} by their spin along the triplet axis, $\vec{S}_{\rm
  TOT}\cdot\hat{e}$.  The transverse components of $\vec{S}_{\rm
  TOT}$, however, do not commute with $\vec\Delta_t$.  An applied
Zeeman field along one of these axes therefore immediately acts to mix
together the former ground and excited states.  It is fairly
straightforward to demonstrate by this mechanism a constant transverse
spin susceptibility for the collinearly-order triplet states.  For
non-collinearly ordered triplets, we conjecture that {\sl all}
components of the uniform susceptibility are finite.  This distinction
is most likely primarily academic, as experimentally available samples
would presumably break up into domains with random orientations of
$\vec\Delta_t$, thus effectively isotropizing the bulk susceptibility.
Very crude estimates for the magnitude of $\chi$ can be obtain in both
the strong and weak coupling limits of excitonically ordered states.
In strong coupling, the susceptibility can be computed by naive
perturbation theory in the mean-field ground state.  In the optimal
case (${\cal H}=0$), in which the excitonic ordering is maximal, one
finds $\chi \sim \mu_{\rm \scriptscriptstyle B}/ {\cal J}_\perp$,
where ${\cal J}_\perp \sim t^2/V$ is the characteristic stiffness for
excitonic ordering (see Sec.~\ref{sec:undoped}), and
$\mu_{\rm\scriptscriptstyle B}$ is the Bohr magneton.  In weak
coupling, the susceptibility is approximately equal to the
free-electron value, $\chi \sim D(\epsilon_F) \mu_{\rm
  \scriptscriptstyle B}$, where $D(\epsilon_F)$ is the density of
states at the Fermi energy.

A puzzling aspect of the experimental data on the hexaborides is the
absence of a substantial gap in optical conductivity measurements in
the undoped materials.  In general, the excitonically ordered
insulators discussed here {\sl are} expected to exhibit hard optical
gaps (i.e. complete absence of weight in $\sigma(\omega)$ at small
$\omega$) at low frequencies and zero temperature, so this is an
important point which such a theory must contend with.  Several
possible physical situations can, however, resolve this apparent
discrepancy.  In the weak and intermediate coupling limits, it is
possible to sustain a metallic state simultaneously with excitonic
order.  This requires imperfectly nested Fermi surfaces -- a detailed
investigation of this possibility is underway.  Even in the strong
coupling limit, it is also possible that a gap exists but is
anomalously small.  Indeed, in the present model, the optical gap can
be estimated by considering the energy cost required to transfer an
electron from one unit cell to its neighbor.  At the optimal
conditions for excitonic ordering, one has $U=2E_{\rm
  \scriptscriptstyle G}+V$ and a straightforward calculation of
energies from Eq.~\ref{tb1}\ gives the optical gap $\Delta_o \approx
V$.  Thus one expects at least $\Delta_o \ll U, E_{\rm
  \scriptscriptstyle G}$.  Note that in the strong-coupling limit,
there is no universal relation between $\Delta_o$ and the excitonic
order parameter.

\subsection{Relevance to the hexaborides}

The models and discussions in this paper demonstrate the feasibility
of a strong-coupling approach to excitonic ordering.  A direct
application of the results to the hexaborides is, however, not
appropriate, due to the simplified nature of the Hamiltonian discussed
here.  It is possible to generalize the tight-binding model discussed
here to a ``two-band'' (p and d orbital) Hamiltonian which more
accurately models the physics of these materials.  This model contains
the significant new ingredient of orbital degeneracy, and hence
considerable additional richness.  Such orbital degeneracy is the
tight-binding analog of the valley degeneracy encountered in
band-theoretic treatments.  The {\sl methods} of this paper, however,
remain applicable in this case as well.  A thorough treatment of this
problem presents an attractive and challenging theoretical
opportunity.

It is reasonable to ask at this point whether there are any
experimental consequences of the excitonic scenario which are
relatively model-independent, and can therefore be firmly stated in
advance of more accurate results?  For this, we look to the discussion
of the previous subsection, focusing particularly on the properties in
the undoped material.  The excitonic scenario postulates symmetry
breaking even without doping, which distinguishes it from, e.g.
low-density ferromagnetism ala Wigner.  All calculations so far appear
to favor triplet ordering, which implies first a {\sl constant
  (temperature independent at low $T$) susceptibility} in the
insulator.  Second, triplet ordering necessarily gives rise to {\sl
  either} static spin moments or static spin currents (or both) within
the unit cell.  Because the latter are presumably difficult to
observe, this is a less strong condition.  Third, because the triplet
state breaks spin-rotational invariance, it implies the existence of
two low-energy collective ``magnon'' modes (presumably dispersing as
$\omega \approx v_s |k|$ at low energies), which could be observable
via inelastic neutron or Raman scattering.  Fourth, an excitonic
explanation for ferromagnetism upon doping requires that the
``pseudo-spin flop'' phenomena occur, and hence (in this sense)
approximate SU(4) symmetry.  This approximate symmetry implies the
existence of additional collective modes with small excitation gaps.

It is natural to ask, given the above emphasis on the undoped state,
whether the excitonic ferromagnet is itself truly a distinct phase of
matter separate from the more familiar (theoretically) Wigner
ferromagnet?  The answer depends upon the extent to and manner in
which the dopant {\sl ions} influence the behavior of the electrons.
In the models investigated to date, the dopants influence the material
only insofar as to donate extra charge carriers, providing no
perturbation to the {\sl potential} felt by the electrons except to
slightly increase the neutralizing positive background charge.  In
this treatment, the lattice point group symmetries are strictly
maintained, and the excitonic ferromagnet is indeed a distinct state
of matter: it exhibits more broken (point group) symmetries than the
Wigner ferromagnet.  In reality, the dopant ions most likely
distribute randomly throughout the crystal, and thereby perturb the
potential experienced by the electrons.  This random potential
explicitly breaks the lattice invariances, and washes out this sharp
distinction between the excitonic and Wigner ferromagnets.

Whether this effect is of practical importance is unclear.  The large
{\sl increase} of conductivity upon doping suggests that the electrons
are not strongly scattered by the dopant ions.  In any case, the
physics at low electron density is quite subtle.  In particular, the
environment around a nearly isolated dopant atom retains a large
fraction of the symmetries of the pure lattice (e.g. a Lanthanum
dopant replacing strontium preserves cubic point group symmetries
around the lanthanum ion).  Because electrons interact with only one
impurity at a time at low densities, the symmetry of the local
environment is expected to improve the distinction between Wigner and
excitonic ferromagnets.  

Clearly, further predictions are possible within more specific models.
Several authors\cite{BalentsVarma,Barzykin}\ have recently pointed out
the likelihood of phase separation at low electron densities.  This
occurs naturally in the pseudo-spin flop picture, but because it has
already been discussed, we will not dwell on it here.  Probably most
importantly, any excitonically-ordered state by definition breaks the
point-group symmetry of the lattice.  This symmetry-breaking is
directly observable, but unfortunately depends in detail on the way it
occurs.  In particular, many of the triplet states that appear to be
favored have less obvious order parameters, so that more work needs to 
be done to ascertain the appropriate experimental probes.  Further
modeling using the strong coupling approach promises to help resolve
these and other issues.




\section{Acknowledgements}
\par

Thanks to C. M. Varma for stimulating interest in this problem, and to Z.
Fisk for providing copies of experimental data.  This research was
supported by the NSF CAREER program under Grant NSF-DMR-9985255.

\vskip -0.2in



\begin{thebibliography}{99}
\bibitem{largeNnote} A alternative large $N$ analysis, which might be
  expected to be {\sl quantitatively} accurate for such an
  SU(4)-invariant model, is currently under investigation.
\bibitem{differentnote} This is a second qualitative difference 
to emerge from the strong-coupling limit relative to weak coupling,
where a singlet-triplet coexistence is preferred\cite{BalentsVarma}.
\end{thebibliography}
\end{document}